\documentstyle[12pt,epsfig]{article}

\begin{document}


\title{$s\bar{s}$ dominance of the $f_0(980)$ meson}

\author{R. Delbourgo, Dongsheng Liu and
	M.D. Scadron\thanks{Permanent address: 
	Physics Department, University of Arizona, Tucson, Az. 85721 USA.}\\
	School of Physics and Mathematics, University of Tasmania\\
	GPO Box 252-21, Hobart, Tasmania, Australia 7001}

\date{\today}
\maketitle

\begin{abstract}\noindent 
We prove that recent data demonstrates unequivocally that the scalar meson
$f_0(980)$ is mostly composed of $s\bar{s}$ quarks and that the coupling 
of $f_0$ to photons and mesons is in agreement with expectations from the
linear sigma model.
\end{abstract}

\vspace{0.5cm}


\vspace{.5in}

The observation\cite{Aul} of the decay mode $\phi(1020) \rightarrow 
\pi^0\pi^0\gamma$ was reported very recently for the first time. The 
experiment clearly shows that the process is dominated by the $f_0(980)\gamma$ 
channel. This may be contrasted with the decay process $J/\psi \rightarrow
\omega\pi\pi$, which was measured a decade ago\cite{Aug} and where it was
found that the $\omega f_0(980)$ channel was highly suppressed but that the 
$\omega\sigma(500)$ and $\omega f_2(1270)$ channels predominated.
In this letter we wish to give an interpretation of these results and show 
that the $f_0(980)$ scalar meson is mostly composed of strange quarks.

First we recall that, according to quark models, the vector meson 
$\omega(782)$ is 97\% nonstrange (viz. NS=$(u\bar{u}+d\bar{d})/\sqrt{2}$)
and 3\% strange (S=$s\bar{s}$). Conversely, the other vector meson
$\phi(1020)$ has the opposite composition: 3\% NS and 97\% S.
We will show that the above two experiments\cite{Aul,Aug}
strongly suggest that the $f_0(980)$ is largely S.
Independently and for different reasons the phenomenological study of
low-energy $\pi\pi$ scattering\cite{TR} and the phase shift analysis\cite{I}
of $\pi K$ scattering have led to the same conclusion about the 
composition of $f_0(980)$.

The effective coupling constants, extracted from ref. [1], are based upon
an inferred $f_0$ total decay width of 188\hspace{-.02in}\raisebox{.07in}{+48}
                                  \hspace{-.27in}\raisebox{-.07in}{-33} ~MeV:
\begin{equation}
 g^{2\,\,eff}_{f_0\pi^+\pi^-}/4\pi = 0.51 \raisebox{.07in}{+0.13}
                \hspace{-.35in}\raisebox{-.07in}{-0.09}{\rm~GeV}^2
\quad {\rm or} \quad |g^{eff}_{f^0\pi^+\pi^-}| \simeq 2.5\pm 0.3 {\rm ~GeV},
\end{equation}
\begin{equation}
 g^{2\,\,eff}_{f_0K^+K^-}/4\pi = 2.10 \raisebox{.07in}{+0.88}
                \hspace{-.35in}\raisebox{-.07in}{-0.56}{\rm~GeV}^2
\quad {\rm or} \quad |g^{eff}_{f^0K^+K^-}| \simeq 5.1\pm 0.9 {\rm ~GeV}.
\end{equation}
These values roughly follow from a dynamically generated theory of the
SU(3) linear sigma model (L$\sigma$M) Lagrangian\cite{DS} where a scalar
meson nonet pattern is demanded: $\sigma_{NS}(670),\kappa(810),\sigma_S(940),
a_0(984)$. A related description\cite{Jaf} is that the scalar mesons are
composed of 4 quarks ($qq\bar{q}\bar{q}$), which is the description adopted
by ref. [1].

In order to bring the observed\cite{PDG} isoscalar mesons $\sigma(600)$ and 
$f_0(980)$ into the L$\sigma$M picture, we must first consider the NS-S
mixing basis,
\begin{equation}
|\sigma\rangle=\cos\phi_s|\sigma_{NS}\rangle -\sin\phi_s|\sigma_S\rangle,
\quad|f_0\rangle=\sin\phi_s|\sigma_{NS}\rangle+\cos\phi_s|\sigma_S\rangle
\end{equation}
in a manner similar to $\eta-\eta'$ mixing\cite{JS}. For the orthogonal
mixed states $\langle \sigma|f_0\rangle = 0$, we find\cite{DS} from (3),
\begin{equation}
m^2_{\sigma_{NS}}=m^2_\sigma\cos^2\phi_s+m^2_{f_0}\sin^2\phi_s,\quad
m^2_{\sigma_{S}}=m^2_\sigma\sin^2\phi_s+m^2_{f_0}\cos^2\phi_s.
\end{equation}
Inserting the dynamically generated NJL-type masses\cite{DS},
\begin{equation}
 m_{\sigma_{NS}}=2\hat{m} \simeq 670{\rm~MeV},\quad 
 m_{\sigma_S}=2m_s\simeq 940{\rm~MeV},
\end{equation}
along with $m_{f_0}\simeq 980$ MeV, one gets
\begin{equation}
 m_\sigma=[m^2_{\sigma_{NS}}+m^2_{\sigma_S}-m^2_{f_0}]^{1/2}\simeq 610
 {\rm~MeV},\,\phi_s=\arcsin\left[\frac{m^2_{f_0}-m^2_{\sigma_{NS}}}
                                      {m^2_{f_0}-m^2_\sigma}\right]^{1/2}
 \!\!\simeq 20^o.
\end{equation}
Such a value for the scalar mixing angle was proposed several years 
ago\cite{JS} and it is worth noting that the proximity in mass between
$\sigma_S(940)$ and $f_0(980)$ has its counterpart in the vector mesons
with $\phi_S(985)$ and $\phi(1020)$. Thus one should not be surprised
that the $f_0(980)$ meson (with $\cos 20^o\simeq 0.94$) is principally an
$s\bar{s}$ state.

In order to link up to the recently extracted interactions (1) and (2), we
first state the predicted L$\sigma$M Lagrangian couplings\cite{DS} (for 
brevity we refer to either the $\pi^+\pi^-$ or $K^+K^-$ final states
as $\pi\pi$ or $KK$),
\begin{eqnarray}
 g_{\pi\sigma_{NS}\pi} & = & (m^2_{\sigma_{NS}}-m_\pi^2)/2f_\pi
   \simeq  2.3 {\rm~GeV}, \nonumber\\
 g_{K\sigma_{NS}K} & = & (m^2_{\sigma_{NS}}-m_K^2)/4f_K \simeq 
   0.45 {\rm~GeV}, \nonumber \\
 g_{K\sigma_SK} & = & (m^2_{\sigma_S}-m_K^2)/2\sqrt{2}f_K \simeq
   2.1 {\rm~GeV},
\end{eqnarray}
where we have substituted the experimental values $f_\pi\simeq 93$ MeV,
$f_K/f_\pi\simeq 1.22$. From these and the mixing angle $\phi_s\simeq 20^o$
we may compute the effective\cite{f1} L$\sigma$M couplings,
\begin{equation}
 g^{eff}_{f_0\pi\pi}=2\sin\phi_s g_{\pi\sigma_{NS}\pi}\simeq 1.6{\rm~GeV},
\end{equation}
\begin{equation}
 g^{eff}_{f_0KK}=2\sin\phi_sg_{K\sigma_{NS}K}+2\cos\phi_sg_{K\sigma_SK}
 \simeq 4.3 {\rm~GeV}.
\end{equation}
Although these predictions (8) and (9) seem shy of (1) and (2) by
64\% and 84\% respectively, we have not yet considered other data.

Specifically, the detailed Particle Data Group (PDG) Tables\cite{PDG} 
quote the average decay rate $f_0\rightarrow\gamma\gamma$ to be 0.56$\pm$0.11
keV. In combination with the $f_0\rightarrow\gamma\gamma$ branching
ratio\cite{PDG} of $(1.19\pm 0.33)\times 10^{-5}$, one may derive a
total $f_0$ width of $47\pm 16$ MeV. Given the $f_0\rightarrow\pi\pi$
branching ratio, one may deduce the PDG effective coupling\cite{f1},
\begin{equation}
 |g^{eff}_{f_0\pi\pi}|_{PDG} = 1.1\pm 0.4{\rm~GeV}.
\end{equation}
($|g_{f_0KK}|_{PDG}$ is unknown, because of the negligible phase space.)

This said, we note that our theoretical values (8) and (9) lie between
the Novosibirsk and PDG values. See Table 1. Regardless
of the final phenomenological resolution of the discrepancy, we must
emphasize that the important result is the contrast between the
measured $\pi\pi$ invariant mass spectra of refs. [1] and [2].

\begin{table}[h]
\begin{tabular}{|l|l|l|l|}
\hline
  & New Data & L$\sigma$M Theory & Particle Data Group \\
\hline
$g^{eff}_{f_0\pi\pi}$ & $2.5\pm 0.3$ & 1.6 & $1.1\pm 0.4$\\
$g^{eff}_{f_0KK}$ & $5.1\pm 0.9$ & 4.3 & - \\
\hline
\end{tabular}
\caption{Comparison of effective couplings (in GeV) of $f_0$ to the 
pseudoscalar mesons.}
\end{table}

\newpage

For the reader's convenience, we display the observed invariant $\pi^0\pi^0$
mass spectrum in $\phi\rightarrow\pi^0\pi^0\gamma$ in Figure 1; our quark model
interpretation is given alongside, in Figure 2. The observed $\pi\pi$ mass
spectra for $J/\psi\rightarrow\omega\pi\pi$ are displayed in Figure 3 to
show the stark difference from Figure 1; again we provide the quark model
interpretation alongside, in Figure 4. The resonance bumps $f_0(980)$ in
$\phi\rightarrow\pi^0\pi^0\gamma$ and $\sigma(500)$ in $J/\psi\rightarrow
\omega\pi\pi$ along with the near absence of an $f_0(980)$ bump in
$J/\psi\rightarrow\omega\pi\pi$ and of $\sigma(500)$ in $\phi\rightarrow
\pi^0\pi^0\gamma$ is telling us that $\sigma(500)$ is mostly NS while $f_0$
is mainly S.  This is in consonance with the linear sigma model and a
scalar mixing angle of $\phi_s \sim 20^o$.

However, to stress that this small mixing angle of $\phi_s$ is not the central
issue, we close this letter by considering $f_0\rightarrow
2\gamma$ decays which proceed via a quark loop. We may compare this radiative
channel with $\pi^0\rightarrow\gamma\gamma$, which is accurately estimated by 
a nonstrange quark triangle; the latter provides the effective amplitude
$\alpha N_c/3\pi f_\pi \simeq 0.025$ GeV$^{-1}$ and agrees with the data
to within 2\% for $N_c=3,\,f_\pi \simeq 93$ MeV.

If $f_0$ were purely nonstrange too, the (isoscalar) $f_0\gamma\gamma$
effective amplitude would be given by
\begin{equation}
 |M(f_{0NS}\rightarrow 2\gamma)| = 5\alpha N_c/9\pi f_\pi\simeq 0.042
 {\rm~GeV}^{-1},
\end{equation}
predicting a decay rate
\begin{equation}
 \Gamma(f_{0NS}\rightarrow 2\gamma)=m_{f_0}^3|M_{NS}|^2/64\pi\sim 8{\rm~keV}.
\end{equation}
On the other hand, if $f_0$ were a pure $s\bar{s}$ scalar, its effective
$\gamma\gamma$ amplitude would be
\begin{equation}
 |M(f_{0S}\rightarrow 2\gamma)| = \alpha N_cg_{f_0ss}/9\pi m_s
 \sim 0.0081 {\rm~GeV}^{-1},
\end{equation}
for a constituent mass $m_s\simeq 490$ MeV and a strange $f_0ss$
coupling\cite{DS} of $\sqrt{2}g_{\pi^0 qq}=\sqrt{2}\,2\pi/\sqrt{3}$. The 
decay rate would then be
\begin{equation}
\Gamma(f_{0S}\rightarrow 2\gamma)=m^3_{f_0}|M_S|^2/64\pi\sim 0.3 {\rm~keV}.
\end{equation}

From (12) we see that the pure  $f_{0NS}\rightarrow\gamma\gamma$ decay
rate is about 15 times larger than the PDG\cite{PDG} average 
rate of 0.56 keV, while the pure $f_{0S}\rightarrow \gamma\gamma$ rate (14)
is within striking range of the experimental value. The comparison can be
improved by including the small mixing angle $\phi_s$. But in any case the
quark triangle description of radiative meson decays reinforces the fact
that $f_0(980)$ is mostly $s\bar{s}$ rather than a ($u\bar{u}+d\bar{d}$)
scalar meson.
\vspace{.3in}

\noindent {\bf Acknowledgements.}

This research was partially supported by the Australian Research Council.
M.D.S. appreciates the hospitality of the University of Tasmania where
this work was carried out.

\vspace{.3in}

\noindent
{\bf Figure Captions}
\vspace{.2in}

\noindent Fig. 1. The measured $\pi^0\pi^0$ invariant spectrum in
$\phi\rightarrow\pi^0\pi^0\gamma$. Reprinted from hep-ex/9807016
by kind persmission from Budker Institute.

\noindent Fig. 2. Theoretical interpretation of $\phi\rightarrow\gamma 
f_0\rightarrow\gamma\pi^0\pi^0$, as due to three gluon exchange.

\noindent Fig. 3. Fit of the $\pi\pi$ distribution in $J/\psi\rightarrow
\omega\pi\pi$, as obtained by the DM2 group, by
kind permission from Dr A Calcaterra.
Reprinted from Nucl. Phys. {\bf B320}, 1, Copyright (1989),
with permission from Elsevier Science.

\noindent Fig. 4. Theoretical interpretation of $J/\psi\rightarrow
\omega\sigma\rightarrow\omega\pi\pi$.

\newpage

\begin{figure}[b!] 
\centerline{\epsfig{file=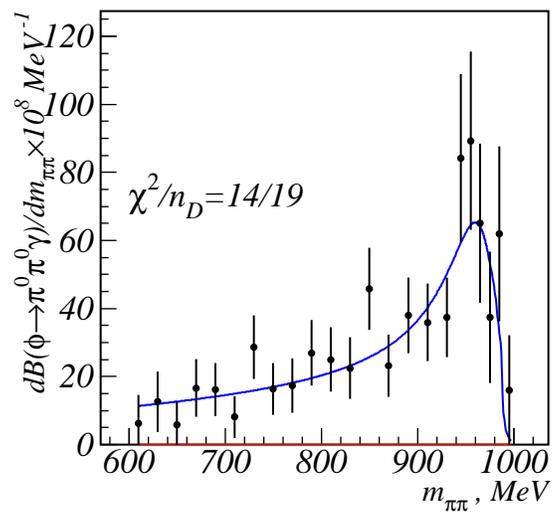}}
\vspace{1pt}
\caption{Delbourgo}
\label{fig1}
\end{figure}

\newpage

\begin{figure}[b!] 
\centerline{\epsfig{file=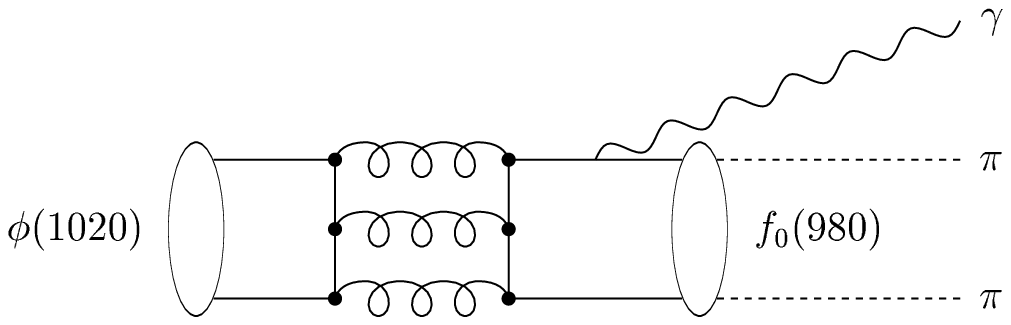}}
\vspace{1pt}
\caption{Delbourgo}
\label{fig2}
\end{figure}

\newpage

\begin{figure}[b!] 
\centerline{\epsfig{file=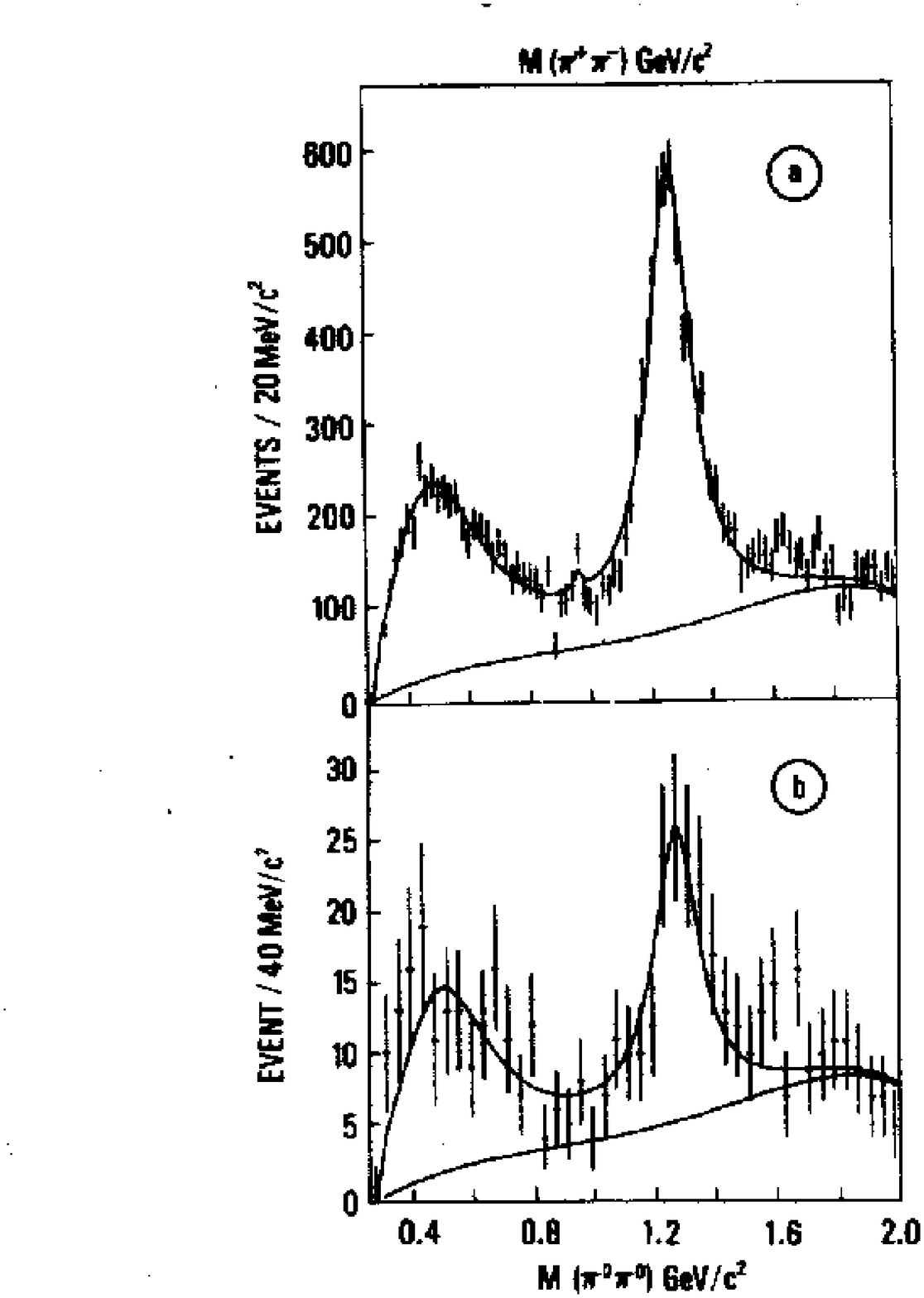}}
\vspace{1pt}
\caption{Delbourgo}
\label{fig3}
\end{figure}

\newpage

\begin{figure}[b!] 
\centerline{\epsfig{file=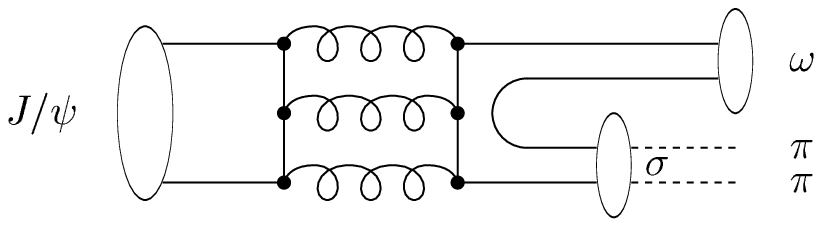}}
\vspace{1pt}
\caption{Delbourgo}
\label{fig4}
\end{figure}


\begin{thebibliography}{99}
\bibitem{f1} Effective couplings are defined via $\Gamma_{f_0\pi\pi} = 
3p|g^{eff}_{f_0\pi\pi}|^2/16\pi m^2_{f_0}$ and
they yield $g^{eff}_{f_0\pi\pi} \simeq 1.12$ GeV for 
$\Gamma_{f_0\pi\pi}=0.781\times 47$ MeV. We need to double the 
Lagrangian couplings by a factor of two to compare them with $g^{eff}$.
\end{thebibliography}
\end{document}